# Slow light enhanced correlated photon pair generation in photonic-crystal coupled-resonator optical waveguides

**Nobuyuki Matsuda[1,2], Hiroki Takesue[1], Kaoru Shimizu[1], Yasuhiro Tokura[1,†], Eiichi Kuramochi[1,2], Masaya Notomi[1,2]**

[1]NTT Basic Research Laboratories, NTT Corporation, Atsugi, Kanagawa 243-0198, Japan

[2]Nanophotonics Center, NTT Corporation, Atsugi, Kanagawa 243-0198, Japan

*m.nobuyuki@lab.ntt.co.jp

[†]Present address: Graduate School of Pure and Applied Science, University of Tsukuba, Tsukuba, Ibaraki 305-8571, Japan

## Abstract

We demonstrate the generation of quantum-correlated photon pairs from a Si photonic-crystal coupled-resonator optical waveguide. A slow-light supermode realized by the collective resonance of high-$Q$ and small-mode-volume photonic-crystal cavities successfully enhanced the efficiency of the spontaneous four-wave mixing process. The generation rate of photon pairs was improved by two orders of magnitude compared with that of a photonic-crystal line defect waveguide without a slow-light effect.

## 1. Introduction

Integrating quantum photonic devices on a small chip [1-17] is under intense study with a view to achieving quantum communication and computation technologies, which have the potential to outperform classical information processing [18-22]. For this purpose, it is crucial to develop sources of non-classical states of light such as single photons and correlated photon pairs on an integrated photonic circuit platform [8-17]. To perform protocols that handle a large number of photonic qubits simultaneously [19-21], we require many independent sources each of which must guarantee stable operation in a simple setup preferably at room temperature. In this regard, photon pair sources based on second- or third-order nonlinear parametric processes in a nonlinear crystal or a waveguide are widely used [8-17, 23-26].

A third-order parametric process, namely spontaneous four-wave mixing (FWM), enables us to obtain a pair of signal and idler photons following the annihilation of two pump photons, where the wavelengths of all the involved photons are closer to each other than in a second-order parametric process [8-17, 25, 26]. Hence, by setting those wavelengths in a telecom band, the source can be made compact by employing integrated nonlinear waveguides developed for telecommunications applications. Si-core integrated waveguides are attracting a lot of attention, because they can be fabricated on a silicon-on-insulator substrate. Centimeters-long Si-wire waveguides have already been proved capable of generating telecom-band photon pairs with sufficient brightness (more than 0.1 pairs/pump pulse) at a non-intense pump power [11]. This is because of the strong confinement of light in the waveguide core area, which is smaller than the size of the wavelength, and the high nonlinearity of Si itself in the telecom band. In addition, Si waveguides can generate photon pairs with a much lower background noise even at room temperature [8-11, 13, 15, 17] than conventional waveguides such as optical fibers [25, 26]. This is because noise photons generated via the Raman scattering process in crystalline Si exhibit a sharp linewidth with detuning of as much as 15.6 THz from the pump frequency and so can be easily eliminated with a wavelength filter.

In addition to noise suppression, downsizing the photon pair source is also of particular importance as regards quantum information applications. This is because spontaneous parametric processes generate photon pairs in a probabilistic way. Hence, attempts to increase the photon pair generation rate will suffer from higher-order events leading to the generation of multiple photon pairs, which will induce errors in subsequent quantum processes. A lower pair generation rate can reduce such higher order events but at the expense of source brightness. A multiplexed single photon source [27] was proposed to solve this problem. The source is a bundle of many identical heralded single photon sources each of which generates photon pairs at a low rate (~ 0.01 pairs/pulse). The whole system provides single photons at a certain period while the multiple pair generation rate is suppressed. However, this scheme requires the connection of many independent photon pair units by high-speed single photon routing technology. Although there was a small-scale implementation very recently [28], further downsizing of the photon pair units is necessary if we are to implement the entire multiplexed system.

A simple solution to the problem of downsizing is to strengthen the optical nonlinearity per unit size. Slowing down the propagation speed of light (*i.e.* the group velocity) in a medium is a significant way of enhancing device nonlinearity, since the slow-light mode can simply prolong the light-matter interaction time. Moreover, the slow-light mode compresses the optical field longitudinally such that its peak intensity increases, which leads to enhanced nonlinearity [29-32]. Slow-light enhancement has been investigated particularly in relation to a Si photonics platform, where sophisticated nanofabrication technologies allow us to engineer the group velocity of light. Slow-light modes in Si photonic-crystal (PhC) line defect waveguides and coupled-resonator optical waveguides (CROWs) [33] based on Si microring resonators have already been shown to enhance the efficiency of spontaneous FWM [13, 15]. On the other hand, using classical stimulated FWM experiments we achieved the highest third-order nonlinearity in Si-core waveguides by employing a CROW consisting of Si PhC nanocavities based on mode-gap confinement [34]. This is attributed to the ultrahigh $Q$ value and the wavelength-sized mode volume of the mode-gap cavities [35, 36]. Hence, we can expect further downsizing of the photon pair source by means of our CROW.

In this work, we report the first generation of correlated photon pairs using a PhC-cavity-based CROW. We enhanced the photon pair generation rate by two orders of magnitude using the slow-light mode of a PhC CROW. We also demonstrate the non-classical correlation of generated photon pairs by showing a violation of the Zou-Wang-Mandel inequality [37], which is valid only for the classical state of light.

## 2. Device

Figure 1(a) is a schematic of our CROW, which we fabricated on a two-dimensional Si PhC slab with a triangular lattice of air holes by electron-beam lithography and dry etching [34-36]. The lattice constant $a$ is 420 nm, the hole radius is $0.25a$ and the thickness of the PhC slab is $0.5a$. Each cavity is formed by the local width modulation of a barrier line defect with a width of $0.98a\sqrt{3}$ (W0.98). The red and green holes are displaced by 8 and 4 nm, respectively, in plane towards the outside. The cavity number is 200, thus the total CROW length is 420 μm. The cavity pitch is $5a$, which yields a supermode that is the collective resonance of all the cavities [34]. Access waveguides were fabricated for the connection between the CROW and external tapered optical fibers. The access waveguides were W1.05 line-defect PhC waveguides (shown in the figure as a region surrounded by purple holes) and Si wire waveguides (not shown). We also fabricated a reference W1.05 waveguide without the CROW section (Fig. 1(b)). Tapered optical fibers were used to couple the light into the waveguides. Fig. 1(c) shows the linear transmission spectrum of the CROW and the reference waveguide measured for TE polarization. The CROW spectrum exhibits a clear transmission band with a width of approximately 6 nm, which represents the supermode formed by the collective resonance of the PhC cavities [33]. The formation can be explained by the tight-binding approximation, and is analogous to the formation of an electronic band in a periodic potential. Due to the ultrahigh $Q$ values ($\sim 10^6$) and small mode volumes of the individual PhC cavities, an optical field can propagate through the CROW without suffering any significant transmission loss, even for a large cavity number [35]. The in- and out-coupling efficiencies $\eta_{\text{couple}}$ and linear loss coefficient $\alpha_{\text{dB}}$ of the reference waveguide are approximately − 8 dB and 2 dB/mm, respectively [34]. For the CROW, we estimate $\eta_{\text{couple}}$ to be − 9 dB, and we also use 2 dB/mm for $\alpha_{\text{dB}}$ taking account of the CROW's peak transmittance of approximately − 20 dB. The slightly smaller $\eta_{\text{couple}}$ value of the CROW is due to the additional waveguide connections between the CROW and the access W1.05 PhC waveguides.

We have already evaluated the $\chi^{(3)}$ nonlinearity of the CROW in classical stimulated FWM experiments [34]. The nonlinearity strength of waveguides that include the slow light effect is represented by the effective nonlinear constant $\gamma_{\text{eff}}$ (/W/m), which scales as $S^2/A_{\text{eff}}$. Here $A_{\text{eff}}$ is the effective lateral mode area, and $S$ is the slowdown factor defined as $S = n_g/n$ [29], where $n_g$ and $n$ are the group and refractive indexes of the waveguide, respectively. The estimated $\gamma_{\text{eff}}$ values for our CROW were 7,200 /W/m at $n_g = 36$ and 13,000 /W/m at $n_g = 49$; these are the largest values yet reported for any nonlinear waveguide with a Si core at a similar $n_g$. This is because a small $A_{\text{eff}}$ is compatible with a large $n_g$ as a result of the strong optical field confinement achieved by the PhC mode-gap nanocavities. Since FWM efficiency per mode is proportional to $\gamma^2$ as we describe later, we can expect a significant enhancement of the photon pair generation rate in the CROW.

## 3. Experiment

Figure 2 shows the experimental setup. The intensity modulator (IM) consists of an electro-absorption modulator and a LiNbO$_3$ intensity modulator, which modulates a continuous beam from the light source operating at a wavelength $\lambda_p$ of 1545.35 nm into a train of pump pulses with a temporal full-width at half maximum (FWHM) $\Delta t$ of 130 ps and a repetition rate $R$ of 100 MHz. The pulses are amplified by an erbium-doped fiber amplifier (EDFA), filtered to eliminate the amplified spontaneous emission noise by the band-pass filter BPFp (3-dB bandwidth: 0.1 nm), and then launched into the waveguides. The $n_g$ values of the CROW and the reference waveguide are approximately 38 and 5, respectively [34]. The output optical fields from the waveguides including correlated photons are collected by another lensed fiber. Then, the light is introduced into notch filters consisting of two fiber-Bragg gratings with a suppression bandwidth of 1.0 nm. Subsequently, an arrayed-waveguide grating (AWG) separates the signal and idler photons into different fiber channels with bandwidths of 0.2 nm (25 GHz). Here we selected the pump-to-signal (or pump-to-idler) detuning $\Delta\lambda_{ps}$ as 0.8 nm (100 GHz), since the half bandwidth of the FWM gain (determined by the FWM phase-matching condition) in the CROW was measured at approximately 1 nm (in a half width) when the pump wavelength was at around the band center (see Ref. 34 for more detail). Then, the photons were passed through BPFs and BPFi for further pump suppression and finally received by InGaAs single photon counting modules (SPCMs) that operated at a gate frequency of 100 MHz synchronized with the pump repetition rate. The quantum efficiency $\eta_{QE}$, gate width, dark count rate, and dead time of the detectors were 17 %, 0.8 ns, 7 x $10^{-6}$ /gate, and 10 μs, respectively. The overall transmittance of the filtration system $\eta_f$ was approximately − 6 dB for both the signal and idler bands, while the transmittance at $\lambda_p$ was less than − 130 dB. The raw coincidence rate $D_c$ (including the accidental coincidence count) and the raw accidental coincidence rate $D_{c,a}$ were determined by measuring the time correlation of the output signals from the two SPCMs using a time-interval analyzer (TIA).

Figure 3(a) is the net photon pair generation rate at the output ends of the waveguides as a function of the coupled average power $P$ of the pump pulses. Here we estimated the net photon pair rate $\mu_c$ by $\mu_c = (D_c - D_{c,a})/(R\,(\eta_{couple}\,\eta_f\,\eta_{QE}\,\eta_{gate})^2)$, where $\eta_{gate}$ is the ratio of the active gates to the 100 MHz clock rate [38]. The number of active detector gates decreases due to the finite detector dead time set in our experiment. The measurement time was 120 s for each data point for good statistics. Fig. 3(a) clearly shows that there is a two-order improvement in the photon pair generation rate when using our PhC-cavity-based CROW compared with the result obtained with the PhC W1.05 waveguide. Thus, we observed the clear slow-light enhancement of photon pair generation from the CROW. Note that the waveguide length $L$ of the CROW is almost a half that of the reference waveguide. The slightly reduced pair generation rate of the CROW at high excitation power is presumed to originate from the free-carrier absorption effect. The free carriers are generated by two-photon absorption in Si. The effect can be avoided using a slab material of a wider band gap such as GaInP [39].

Under the slowly varying envelope approximation we can obtain $\mu_c = \Delta\nu \Delta t \, (\gamma_{\text{eff}} P_{\text{peak}} L_{\text{eff}})^2$ when $\Delta\lambda_{\text{ps}}$ is approximately less than the half-bandwidth of the FWM gain, where $\Delta\nu$ is the bandwidth of the signal and idler photons (determined by the channel bandwidth of the AWG). $P_{\text{peak}} = P/(R \Delta t)$ is the coupled pump peak power. $L_{\text{eff}}$ is the effective waveguide length associated with $L_{\text{eff}} = (1 - \exp(-\alpha L))/\alpha$, where $\alpha$ is the attenuation coefficient on a linear scale. We see that all the data in Fig. 3(a) exhibited good $P^2$ dependence. From the fitted function of the CROW (represented by the dashed black line) we obtained a $\gamma_{\text{eff}}$ of 9,000 /W/m. The $\gamma_{\text{eff}}$ value is in good agreement with that of our previous classical stimulated FWM experiments [34]. In addition, the value is again the highest for any Si-core nonlinear waveguide yet reported. Meanwhile, $\gamma_{\text{eff}}$ was 650 /W/m for the reference waveguide. Although the ratio of their squared slow-down factor $S^2$ is $(38/5)^2 \sim 58$, the $\gamma_{\text{eff}}$ ratio of the two waveguides is 14. One reason for the discrepancy between the ratios is that the photon pair generation rate from the reference waveguide includes pairs generated in the access Si-wire waveguides. Hence, the net pair generation rate in the reference waveguide would be lower than observed. Another reason is that we possibly underestimated the $\gamma_{\text{eff}}$ of the CROW by using an $\alpha_{\text{dB}}$ value of 2 dB/mm, which was the value for the W1.05 PhC waveguide. If we assume a scaling of $\alpha_{\text{dB}} \propto S$, we estimate the $\alpha_{\text{dB}}$ of our CROW to be $-15.2$ dB/mm, from which we can obtain a $\gamma_{\text{eff}}$ of 15,600 /W/m; a more precise discussion requires an additional rigorous estimation of the waveguide loss. We note that the photon pair generation occurring in 6 m long optical fiber links (the section between the BPFp and the notch filters) was negligible (Fig. 3(a)). The estimated $\gamma$ value for the optical fiber link was $2.1 \times 10^{-3}$ /W/m, which agrees well with previously reported values [40].

Figure 3(b) shows the measured coincidence to accidental coincidence ratio (CAR = $D_c/D_{c,a}$) of the CROW with respect to $P$. CAR > 1 represents the time correlation of photon pairs. The experimental CAR values ranged between 1 and 2, exhibiting the time correlation. Meanwhile, we obtained a maximum CAR of 8 for the reference waveguide. These CAR values are lower than those obtained in many previous spontaneous FWM experiments. To discuss the reason for this, we show the net generation rates of photons in the frequency band for the signal photons $\mu_s = N_s/(R \, (\eta_{\text{couple}} \eta_f \eta_{\text{QE}} \eta_{\text{gate}}))$ in Fig. 3(c), where $N_s$ is the raw single count rate measured by the SPCM set in the signal channel excluding the dark count rate of the detector. In most of the region, $\mu_s$ is dominated by the component that exhibits $P^1$ dependence. This indicates the presence of contaminating photons scattered via processes other than spontaneous FWM. A possible process is the spectral broadening of the pump pulses induced by self-phase modulation, where the amount of nonlinear phase shift is proportional to $P$. There is also a possibility of the occurrence of other inelastic scattering processes [41]. The small pump-to-signal detuning $\Delta\lambda_{ps}$ of 0.8 nm in our experiment (limited by the FWM gain bandwidth) in contrast to that in conventional experiments performed with $\Delta\lambda_{ps} \geq 5$ nm [8-17, 25, 26], means that our experimental setup can suffer significantly from such noise photons. In addition, the CAR value of the CROW was lower than that of the reference waveguide. This is because the spurious linear scattering process can also be enhanced by a factor of $S$.

We estimate CAR values using the expression of CAR = $1 + \eta^2\mu_c/(\eta\mu_s + \mu_d)^2$ [11], where $\eta = \eta_{couple}\,\eta_f\,\eta_{QE}$ is the photon detection efficiency and $\mu_d$ is the dark count rate of the detector for both the signal and idler channels. Here $\mu_s$ is assumed to be the same as the net photon generation rate in the idler channel. For $\mu_c$ and $\mu_s$ we used the fitted functions represented by the dashed curves in Figs. 3(a) and 3(c), respectively. The estimated CAR is shown by the solid curve in Fig. 2(b), which agrees well with the experimental data. The dashed line represents for the case of $\mu_d = 0$, which still remains low. Thus, the smallness of the CAR values in our experiment is mainly caused by the noise photons included in $\mu_s$ described above.

We show the non-classicality of the photon pairs by testing the violation of the Zou-Wang-Mandel inequality $V \leq 0$ for the classical state of light, where $V = (D_c - D_{c,a}) - 2(D_s - D_{s,a} + D_i - D_{i,a})$ [37]. Here $D_s$ ($D_i$) and $D_{s,a}$ ($D_{i,a}$) are the self-correlated coincidence rates and accidental coincidence rates obtained by splitting the signal (idler) port using a 50/50 beam splitter and examining the time correlation between the two beams (for a measurement time of 240 s). The experimental $V$ values of generated photon pairs in a CROW are plotted in Fig. 4 for several pump peak powers. $V > 0$ was obtained in the $0.02 < P < 0.3$ (mW) range, and exhibit non-classicality. We also show the extent of the classicality violation characterized by $V/\sigma$ in the same figure, where $\sigma$ is the standard deviation of the data. A maximum violation as high as 20$\sigma$ was obtained at $P$ = 0.1 mW. Thus, we confirmed the generation of non-classical photon pairs from the PhC-cavity-based CROW.

## 4. Discussion and Conclusion

We succeeded in observing quantum-correlated photon pair generation from a PhC CROW. From the CROW data in Fig. 3(a), the photon pair generation efficiency was estimated to be 1.2 × 10$^9$ pairs/pulse/W$^2$/m$^2$/nm (here m$^2$ refers to the squared length of the device). This is 2,600 times higher than that of a Si wire waveguide [11], suggesting the possibility of downsizing the photon pair source by using a PhC CROW. However, for practical use it is necessary to improve the CAR compared with the observed values. The straightforward way to achieve this is to engineer the dispersion property of the CROW. By increasing the FWM bandwidth of the CROW, we could perform the experiment under a large $\Delta\lambda_{ps}$ condition, by which we can expect to reduce the number of noise photons that leak into the wavelength windows for the signal and idler photons.

In addition, this is the first report of photon pair generation using the intra-band FWM process in CROWs. Very recently, Ong and Mookherjea proposed that the intra-band FWM of a CROW is useful for applications such as tailoring a two-photon joint spectrum [42]. However, such an experiment requires sharp wavelength filters to separate the generated photons as well as the suppression of the pump field inside the Bloch band of a CROW, which is generally narrow. Our experiment shows the possibility of such an experiment, since we realized the separation of photon pairs with a pump-to-signal detuning of only 100 GHz, which is the smallest value yet reported.

In summary, we have demonstrated the generation of quantum-correlated photon pairs from a PhC-cavity-based CROW. The generation rate was improved by two orders of magnitude by using the collectively resonant supermode of PhC nanocavities compared with that of a PhC line-defect waveguide. This result is a step towards the further downsizing of the photon pair source for large-scale quantum optics experiments.

## Acknowledgments

We are grateful to Ken-ichi Harada for fruitful discussions. This work was supported by a Grant-in-Aid for Scientific Research (No. 22360034) from the Japan Society for the Promotion of Science.

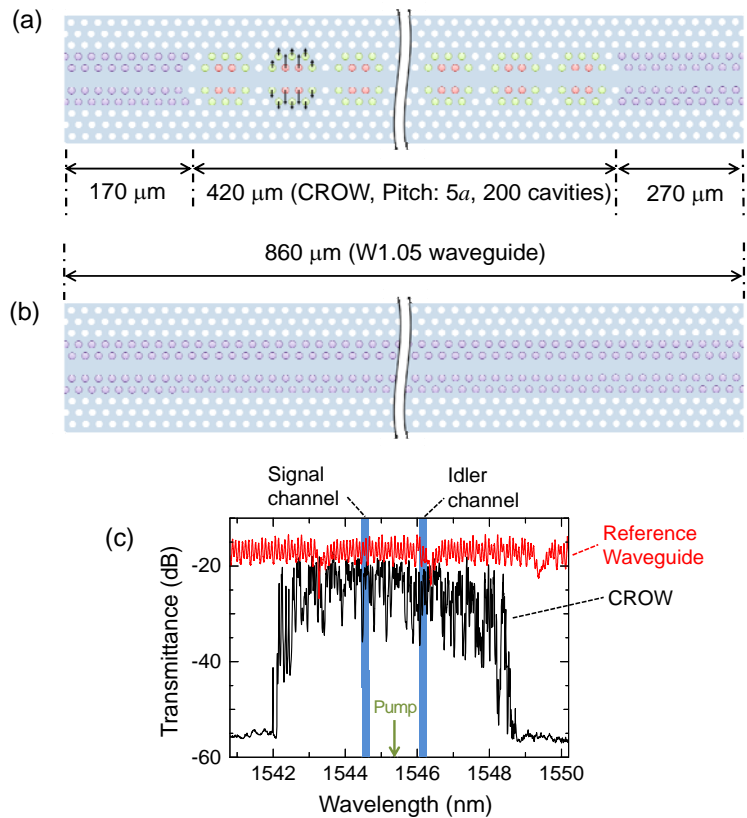

Fig. 1. (a) A CROW consisting of photonic-crystal mode-gap nanocavities (top) and a reference waveguide implemented without a CROW (bottom). (c) Linear transmission spectra of the CROW and the reference waveguide. We also show the center wavelength of the pump pulses and the transmission window of the wavelength-division-multiplexing filter used for the spontaneous FWM experiment (see section 3).

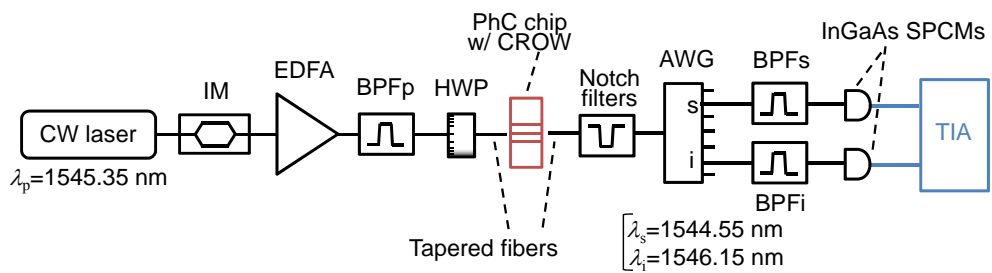

Fig. 2. Experimental setup. IM: intensity modulator, EDFA: erbium-doped fiber amplifier, BPF: band-pass filter, HWP: half-wave plate, AWG: arrayed-waveguide grating, SPCM: single-photon counting module, TIA: time-interval analyzer.

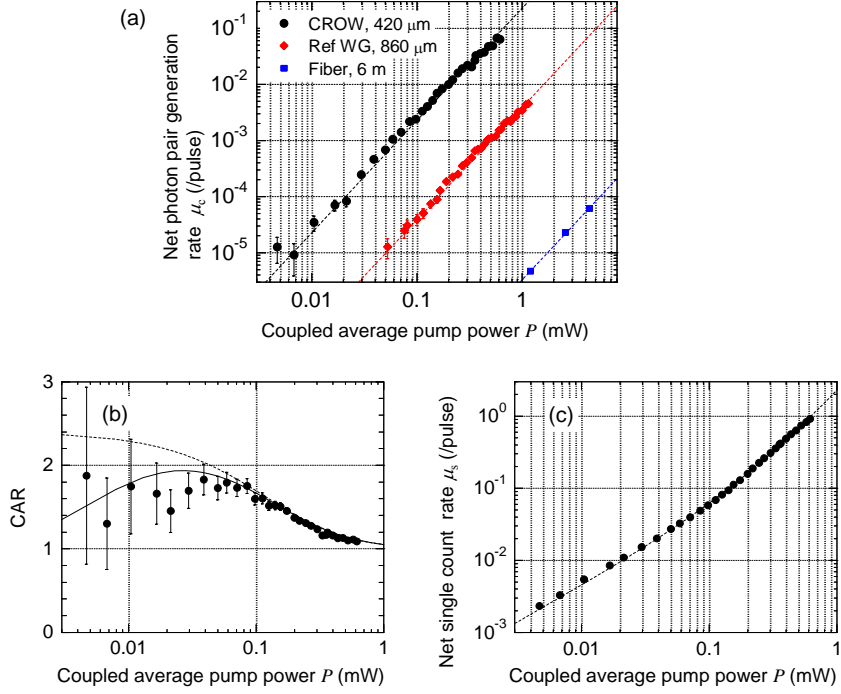

Fig. 3. (a) Net photon generation rate from various waveguides as a function of the coupled average pump power $P$. Dashed lines represent fitting results by $P^2$. (b) CAR values and (c) the net photon generation rate inside the wavelength band for the signal photons versus $P$. In (b), the solid and dashed curves are estimated CAR values (calculated in accordance with the procedure in Ref. 11) with and without the dark count rate, respectively. In (c), the dashed curve shows second-order polynomial fitting.

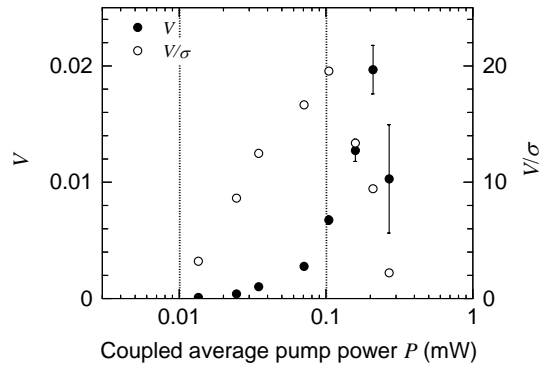

Fig. 4. Zou-Wang-Mandel parameter $V$ and violation extent of classicality $V/\sigma$, as a function of $P$.